# Multiparametric Deep Learning Tissue Signatures for a Radiological Biomarker of Breast Cancer: Preliminary Results

Vishwa S. Parekh, MSE[1,3], Katarzyna J. Macura, MD, Ph.D.[1,2], Susan Harvey MD,[1]

Ihab Kamel MD, Ph.D.[1,2], Riham El-Khouli, MD, Ph.D.,[1,4]

David A. Bluemke, MD, Ph.D[5]., Michael A. Jacobs, Ph.D.[1,2]

[1]The Russell H. Morgan Department of Radiology and Radiological Sciences and
[2]Sidney Kimmel Comprehensive Cancer Center.
The Johns Hopkins University School of Medicine, Baltimore, MD 21205, USA
[3]Department of Computer Science,
The Johns Hopkins University, Baltimore, MD 21208
[4] Department of Radiology and Radiological Sciences,
University of Kentucky, Lexington KY 40536
[5] Department of Radiology,
University of Wisconsin School of Medicine and Public Health.
Madison, WI 53726.

Address Correspondence To:
Michael A. Jacobs
The Russell H. Morgan Department of Radiology and Radiological Science and Oncology,
Division of Cancer Imaging
The Johns Hopkins University School of Medicine,
Traylor Blg, Rm 309
712 Rutland Ave, Baltimore, MD 21205
Tel:410-955-7483
Fax:410-614-1948
email: mikej@mri.jhu.edu

Keywords: Deep Learning, Machine learning, Magnetic resonance imaging, multiparametric MRI, Breast, Cancer, Diffusion, tissue biomarkers.

*Preprint before review*

**Advances in knowledge:**
1. Advanced computational methods using multiparametric deep learning (MPDL) with multiparametric MRI are significant predictors of malignant or benign breast lesions.

2. Development of a "tissue based" deep learning method validated with an independent radiological data set.

3. The MPDL model with pharmacokinetic modeling parameters and diffusion weighted imaging/Apparent Diffusion Coefficient metrics demonstrated similar diagnostic performance of the radiologist in characterizing breast lesions.

**Implications for patient care**: The integration of advanced computational techniques and artificial intelligence methods to assist radiologists will become available in the future reading rooms and will transform medicine in general. Deep learning methods will be the conduit for modeling of clinical and radiological variables which will provide the foundation for radiological precision medicine in patients.

**Abstract**

**Background and Purpose:** Integration of advanced computational techniques and artificial intelligence methods to assist radiologists will become available in future reading rooms and will transform medicine in general. This study was conducted to evaluate the feasibility and role of a novel deep learning method using multiparametric breast Magnetic Resonance Imaging(mpMRI) and defining tissue signatures for improved automated detection and characterization of breast lesions.

**Methods:** We developed and tested a multiparametric deep learning(MPDL) network for segmentation and classification of breast MRI in 171 patients. The MPDL network was constructed from stacked sparse autoencoders. MPDL network inputs were T1 and T2-weighted imaging, diffusion weighted imaging(DWI) and ADC mapping, and dynamic contrast enhanced(DCE) imaging tissue signatures. Evaluation of MPDL consisted of cross-validation, sensitivity and specificity. Dice similarity between MPDL and post-DCE lesions were evaluated. Statistical significance was set at $p≤0.05$.

**Results:** The performance of MPDL on Modified National Institute of Standards and Technology database(MISNT) data set was 99.7%. For MRI validation set, a 4.2%±3.6% percent difference between volumes was found between MPDL and test data set. The MPDL segmented glandular, fatty, and lesion tissue with an overlap of 0.87±0.05 for malignant patients and 0.85±0.07 for benign patients. The sensitivity and specificity for differentiation of malignant from benign lesions were 90% and 85% respectively with an AUC of 0.93

**Conclusion:** Integrated MPDL method accurately segmented and classified different breast tissue from multiparametric breast MRI. Deep learning can be used to construct a personalized database of tissue signatures with accurate characterization of different tissue types.

# Introduction

A new paradigm is beginning to emerge in Radiology with the advent of increased computational capabilities and algorithms. This has led to the ability of "real time" learning by computer systems of different lesion types to help the radiologist in defining disease. In particular using deep learning algorithms to segment and classify different radiological images. We chose to use multiparametric magnetic resonance imaging (mpMRI) parameters which capitalize on the different contrasts of tissue. For example, using conventional and advanced MRI parameters of T1- and T2-weighted, diffusion-weighted imaging (DWI) and dynamic contrast enhanced imaging (DCE) provide qualitative and quantitative information of different tissue types which can be used to construct "tissue" signatures information of tissue [1-4]. Therefore, to integrate mpMR and characterize breast tissue, we have developed a machine learning method coupled with deep learning for segmentation and characterization of breast tissue using mpMRI. Deep learning networks (DLN) allow for the "learning" of radiological relationships between different tissue types and provides new methods to "segment" and/or classify high-dimensional data sets[5-11]. These DLN algorithms allow for accurate and reliable prediction of tissue types from "raw" input images with the aim to improve the radiologist's clinical decision support in different diseases[12-15]. Therefore, we implemented an unsupervised deep learning system based on stacked sparse autoencoders (SSAE). Autoencoders are unsupervised neural networks that are trained to create a compact or a low dimensional representation of its input via the hidden layer [7,9,16,17]. The stacked sparse autoencoder network (SSAE) is a stack of sparse autoencoders with each autoencoder forming a layer of the SSAE. This allows us to use deep learning to develop multiparametric breast tissue signatures across subjects, without prior knowledge of the lesion type for application to patients for tissue segmentation and classification of breast lesions.

In this work, we establish the use of a multiparametric deep learning SSAE for radiological biomarkers of breast tissue by demonstrating that the MPDL SSAE can segment different breast tissue types, i.e., fatty, glandular, and lesion tissue. Second, we can classify the breast lesions into benign or malignant and show that the results are similar to radiologists. Finally, we developed and validated the MPDL tissue signature model with an independent data set in breast cancer patients.

# Materials and Methods

**Clinical subjects**: All studies were performed in accordance with the institutional guidelines for clinical research under a protocol approved by the Johns Hopkins University School of Medicine Institutional Review Board (IRB) and all HIPAA agreements were followed for this retrospective study. One hundred and eighty-nine women (96 malignant, 39 benign and four normal (no-lesion) and were scanned. Malignancy was determined by pathology in all cases. Fifty (n=50) cases were obtained from University California San Francisco (UCSF) for an independent deidentified test data set. These fifty cases are from a Phase 3 clinical trial for women receiving neoadjuvant chemotherapy for locally advanced breast cancer defined by histology[18,19]. We used the baseline study before initiation of the therapeutic regimen.

**Multiparametric MRI imaging protocol**: MRI scans were performed on a 3T magnet (Philips North America Corporation), using a dedicated phased array breast coil with the patient lying prone with the breast in a holder to reduce motion. MRI sequences consisted of fat suppressed (FS) T2WI spin echo (TR/TE=5700/102) and fast spoiled gradient echo (FSPGR) T1WI (TR/TE =200/4.4, Field of View (FOV)=256x256, slice thickness(ST), 4mm, 1mm gap); diffusion-weighted (TR/TE=5000/90ms,b=0-800, FOV= 192x192,ST=6mm); and finally, pre- and post-contrast enhanced images FSPGR T1WI (TR/TE=20/4, FOV=512x512, ST=3 mm) were obtained after intravenous administration of a GdDTPA contrast agent (0.2mL/kg(0.1 mmol/kg)). The contrast agent was injected over 10 seconds, with MRI imaging beginning immediately after completion of the injection and the acquisition of 14 phases. The contrast bolus was followed by a 20cc saline flush. The DCE protocol included two minutes of high temporal resolution (15 sec per acquisition) imaging to capture the wash-in phase of contrast enhancement. A high spatial resolution scan for two minutes then followed, with additional high temporal resolution images (15 sec per acquisition) for an additional two minutes to characterize the wash-out slope of the kinetic curve for pharmacokinetics(PK)[2]. Total scan time for the entire protocol was less than 45 minutes.

The independent validation breast MRI scans were acquired on a different 1.5 T magnets using a dedicated breast RF coils and obtained from the ACRIN I-SPY clinical trial[18,19]. The images used for validation were fat suppressed, T1 weighted dynamic contrast enhanced series obtained unilaterally in the sagittal orientation with TR≤20ms, TE=4.5ms, flip angle≤45°, FOV: 160-180, matrix size > 256x192, ST ≤ 2.5 mm.

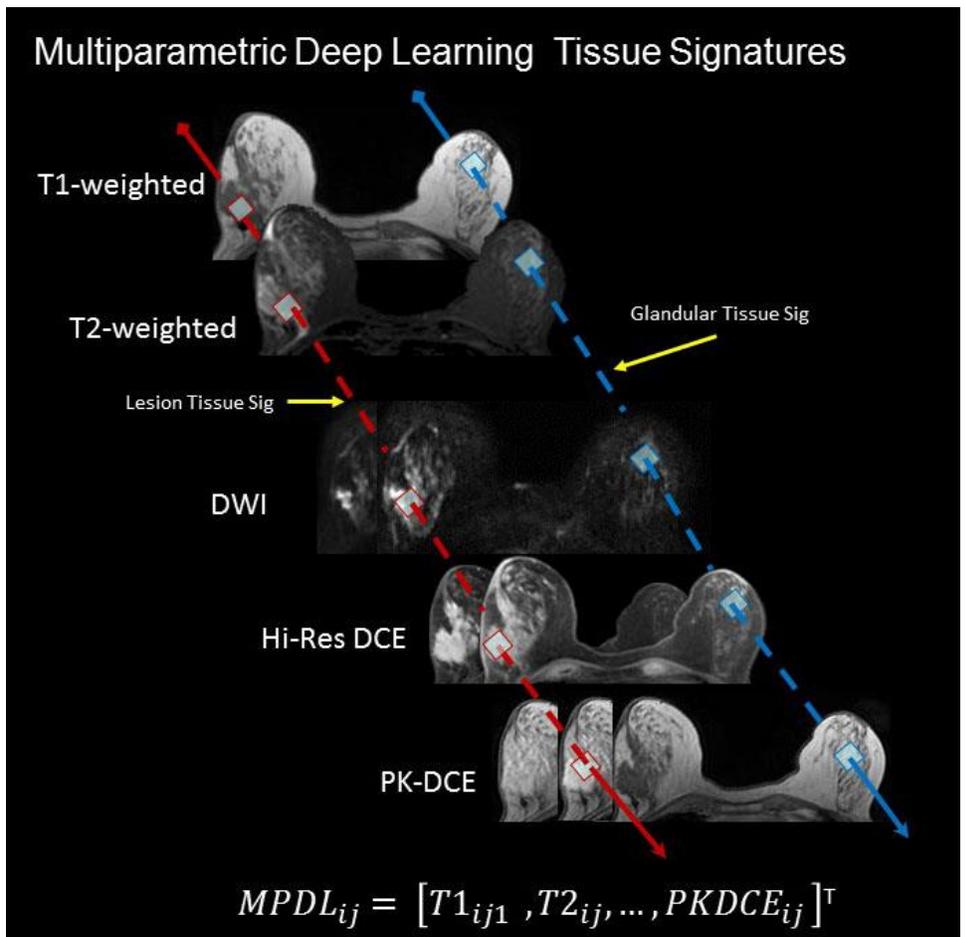

**Figure 1.** *Demonstration of the multiparametric deep learning tissue signatures on axial mpMRI of the breast. Representative tissue signatures from the normal and abnormal tissue are obtained on each of the input MRI to create the vector signature $MPDL_{ij}$.*

**Multiparametric Image Registration:** The mpMRI were coregistered using a hybrid registration algorithm that combines 3D wavelet transformation for 3D reslicing and rescaling of the MRI volumes with nonlinear affine transformation to minimize the loss of information in image transformations [20]. The pre-contrast image of the DCE dataset was used as the reference image for all the other MRI images.

**Multiparametric MRI tissue signature generation:** The Eigenimage filter (EI) segmentation algorithm was used to segment the breast lesions from the post contrast DCE image. The EI is a linear filter that maximizes the projection of a desired tissue (lesion tissue) while it minimizes the projection of undesired tissues (glandular tissue) onto a composite image called an Eigenimage [21,22]. Tissue signatures of glandular, fatty, and lesion tissue were defined using the EI filter are described below. Moreover, the EI corrects for partial volume effects, this allows for better demarcation of the underlying structures [23].

**Multiparametric Deep Learning Tissue Signatures:** The multiparametric deep learning network was trained on the breast tissue signatures defined using EI identified on all the original breast MRI images as demonstrated in **Figure 1**. The MPDL network builds a composite feature representation of the breast tissue signatures of the underlying breast tissue. A voxel tissue signature vector is defined as the vector of gray level intensity values corresponding to that voxel position in each image in the entire data sequence (n=23 images). Mathematically, the MPDL tissue signature is defined as follows:

$$MPDL\ Tissue\ Signature = [T_1, T_2, DWI, \cdots DCE_n]^T$$

For this study, four sets of tissue signature vectors were defined. The first set of tissue signatures for normal tissue, $Normal = [N_1, N_2, \cdots N_n]^T$ was chosen from the glandular tissue, second one for fatty tissue, $Fatty = [F_1, F_2, \cdots F_n]^T$, a third one for lesion tissue, $Lesion = [L_1, L_2, \cdots L_n]^T$ and a fourth one for background noise. Each set of MPDL tissue signature vectors created automatically using a multiparametric region growing algorithm. The initialization to the region growing algorithm is provided by the operator identifying pixels within the tissue of interest. The tolerance for region growing was set at 5%. The final ROI is created by computing a logical AND operation between the ROIs generated from region growing on each of the MR images (**Figure 1**). By using several images concurrently, the probability of a pixel from another tissue being included in the final ROI (due to noise, partial volume, and nonuniformities) is reduced. The computer time required for producing the final ROI was less than a second for each tissue type.

**Multiparametric Deep Learning Network**
**Stacked Sparse Autoencoders Network Architecture:** We developed the mpMRI segmentation deep network by stacking sparse autoencoders (SSAE) for segmenting a multiparametric breast MRI dataset into regions corresponding to different tissue types.

**Figure 2** demonstrates the network architecture of the mpMRI segmentation deep network. Each sparse autoencoder of the SSAE was pre-trained in an unsupervised fashion to create a low dimensional representation of its input via the hidden layer by the tissue signature vectors. The input to each autoencoder except the first autoencoder was the hidden layer representation discovered by the previous layer. The first layer autoencoder learns a low dimensional representation, $Y = \{y^{(1)}, y^{(2)}, ..., y^{(N)}\} \in R^d$ from the training input tissue signatures, $X = \{x^{(1)}, x^{(2)}, ..., x^{(N)}\} \in R^D$, where D is the dimensionality of the input tissue signatures, N is the total number of tissue signatures and d is the number of nodes in the hidden layer of the autoencoder. The output of the final sparse autoencoder was used as input to train a softmax classifier to identify the input tissue signature as background, fat, glandular or lesion.

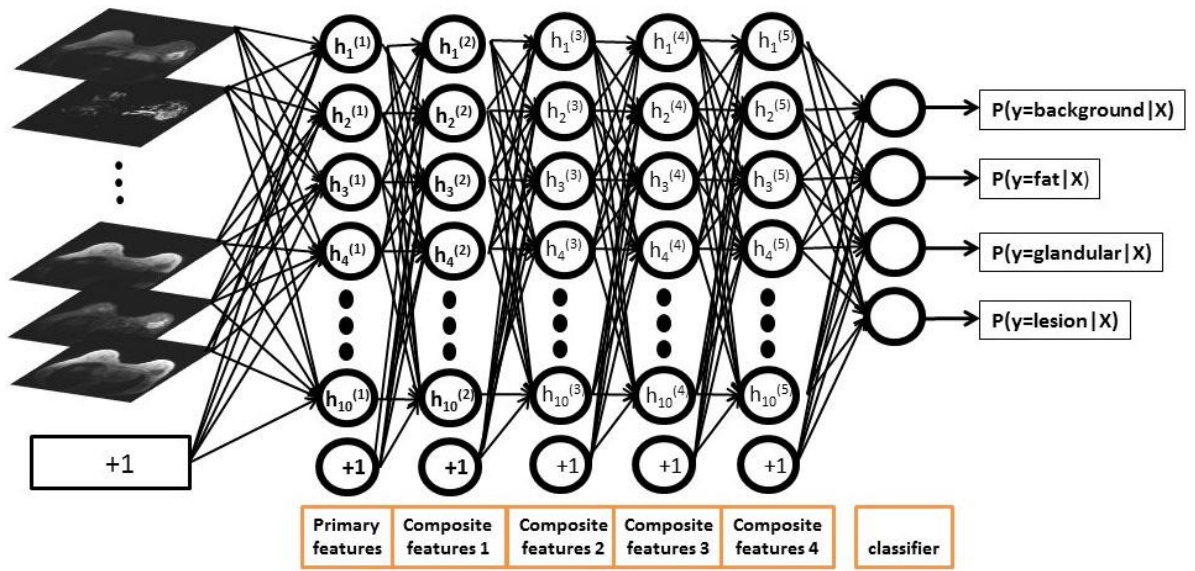

**Figure 2**. *Framework for mpMRI segmentation deep network trained to segment multiparametric breast MRI into regions of different tissue types and background. The stacked sparse encoder deep network is constructed of ten hidden layers with five nodes each and a softmax classification layer that outputs the probability of different tissue types for the input tissue signature.*

**Multiparametric MRI segmentation deep network training and evaluation:** For the MpMRI segmentation, the training parameters obtained from optimization were set as follows: the number of

layers=5, the number of nodes in each hidden layer=10, L2 regularization penalty = 0.001, Sparsity proportion = 0.5, Sparsity regularization = 4. The transfer function for the autoencoder nodes was selected as the saturating linear function given as

$$f(x) = \begin{cases} 0, if\ x \leq 0 \\ x, if\ 0 < x < 1 \\ 1, if\ x \geq 1 \end{cases}$$

We tested the MPDL for segmentation of breast tissue into different tissue classes using a two-fold cross validation. The balance between number of tissue signatures used to train different tissue types were maintained by sampling uniformly at random equal number of tissue signatures corresponding to each tissue type from each patient. Moreover, the MPDL tissue signature can adapt to the input sequences into the SSAE.

To perform a quantitative comparison between the MPDL segmented regions, we defined the radiological ground truth for breast imaging by the EI segmented regions. The dice similarity index (DS) was used as the overlap evaluation metric [24]. The dice similarity index is designed to find the similarity between overlapping regions from two objects. Mathematically, DS is given by the following equation:

$$DS = \frac{2(A \cap B)}{n(A) + n(B)}$$

Here, A and B are the lesion areas obtained by ground truth (EI segmented post-contrast image) and the multiparametric deep learned image, respectively. The EI segmentation was obtained by thresholding the EI contrast image. The threshold was obtained by evaluating the post-contrast MR image histogram, and using the mean and a 95% confidence interval.

**Multiparametric MRI classification deep network training and evaluation:** For the MpMRI classification, we developed a hybrid feature extraction and classification method termed SAE-SVM by combining sparse autoencoder (SAE) with support vector machine (SVM) algorithm. The unsupervised SAE component of the SAE-SVM algorithm automatically extracted the intrinsic tissue signatures for each MRI parameter which were then trained by the linear SVM classifier to predict a tumor as benign or malignant. The training parameters of the SAE were set as follows: Number of nodes in the hidden layer=10, L2 regularization penalty=0.001, Sparsity proportion=0.5, Sparsity regularization=4, Encoder transfer function: sigmoid, Decoder transfer function: linear. The imbalance in the number of benign and malignant patients was resolved by setting a higher misclassification cost for benign than malignant patients. The SAE-SVM feature extraction and classification method was tested using leave-one-out and ten-fold cross validation with sensitivity, specificity and area under the receiver operating characteristic curve (AUC) as the evaluation metrics. The imbalance in the number of benign and malignant patients was resolved by setting a higher misclassification cost for benign than malignant patients. The optimal value of the misclassification penalty was obtained using grid search on misclassification penalty ratios from the set $\text{Benign: Malignant} = \{1:1, 1.5:1, 2:1, 2.5:1, 3:1, 3.5:1, 4:1\}$.

**Validation of the Multiparametric Deep Learning Network.**
**Magnetic Resonance Imaging Data:** We used 50 patients with calculated volumes from the University California at San Francisco I-Spy ACRIN study to test our MPDL network[18,19]. To compare with our dataset, we used the baseline DCE contrast imaging session from the study[18,25,26]. The UCSF data was registered to the pre-contrast DCE image. After application of the MPDL, the MPDL and UCSF volumes were compared and analyzed.

**Statistical Methods:** We computed summary statistics (mean and standard deviations) from the quantitative imaging parameters from the mpMRI. The percentage difference and overlap segmented from the lesion areas were computed and compared. Statistical analysis was performed using linear regression to correlate the total lesion areas. A Student's t-test was used to determine statistical significance between the lesion areas and the percent difference between the lesion boundaries. Sample size was calculated based on the ROC curve by combining information from multiple MRI tissue contrasts [27-29]. A sample size of 112 can give 85% power to detect an increase in specificity (under significance level alpha=0.05).

The same sample size also gives us greater than 85% power to differentiate sensitivities between 80% and 95% at alpha=0.05 significance level. We increased our total sample size from 112 to 189. Bland-Altman tests were run to insure no bias in the validation data set. Sensitivity and specificity with Area under Curve (AUC) was computed. Statistical significance was assigned for $p < 0.05$.

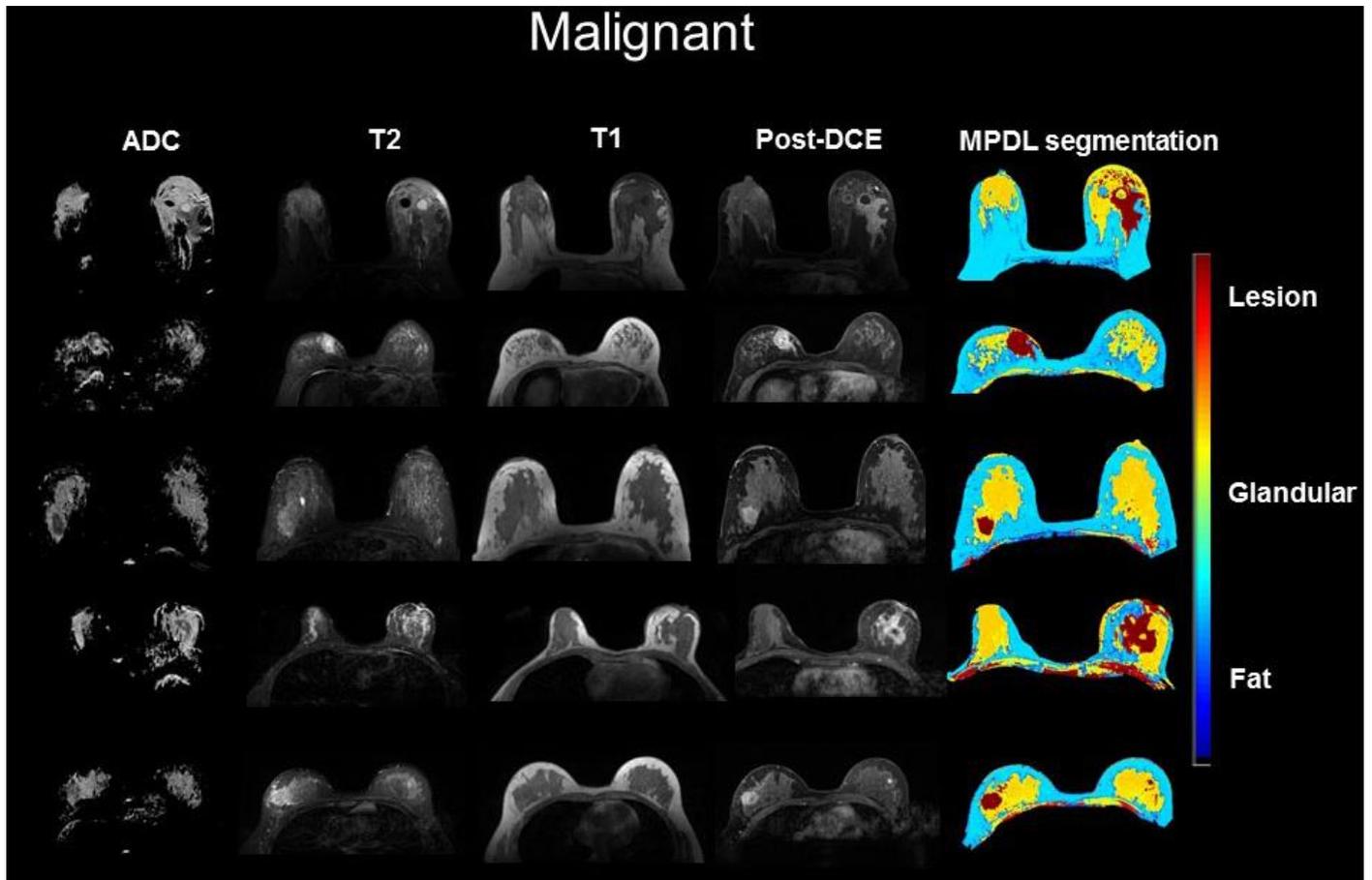

**Figure 3.** *Illustrates the use of MPDL network on axial breast mpMRI in five representative malignant patients. The color coding for different tissue types are shown to the right of the images.*

## Results

**Patient Demographics**: There were total of 139 women with suspicious breast lesions in this study (n=139 training). For the training set (n=139) the mean age was 52 years (range: 24-80 years). Ninety-six patients had malignant breast lesions (69%) while thirty-nine patients had benign breast lesions (28%) and four had no lesions identified (3%).

**Quantitative mpMRI:** In the training data set, the DWI and DCE sequences provided quantitative radiological metrics. there were significant differences (p<0.001) between the ADC map values for malignant and benign breast lesions. ADC values for malignant cases were (mean and standard deviation) 1.26±0.13 (mm$^2$x10$^{-3}$/s) and benign lesions were 1.74±0.17 (mm$^2$x10$^{-3}$/s). Glandular tissue ADC values for malignant and benign lesions were not significantly different, 2.16±0.46 and 2.34±0.33 (mm$^2$x10$^{-3}$/s), respectively. The DCE PK values were significantly different (p<0.05) between malignant and benign lesions. The $K^{trans}$ values were 0.55±0.32 (1/min) and EVF were 0.30±0.16 for malignant cases and 0.25±0.19 (1/min) and 0.22±0.13 for benign cases, respectively.

**Training Data Set:** The MPDL tissue signatures were defined for different breast tissue types **(Figure 1)** and applied to the 139 mpMRI breast cases. **Figure 2** demonstrates the mpMRI deep network segmentation for the tissue signatures. **Figure 3** illustrates the results on five representative malignant patients. Similarly, **figure 4** illustrates the mpMRI deep network segmentation results on five benign patients. **Figures 3 and 4**, demonstrate the mpMRI deep network successfully segmented different tissue types from benign and malignant patients using the tissue signature model. The dice similarity index between the lesion segmentations demonstrated excellent overlap with mean and standard deviation (SD) 0.87±0.05 for malignant patients and 0.85±0.07 for benign patients. Representative cases are shown in **Figure 5.** The optimal value of misclassification ratio was obtained at 2:1 i.e. benign patients had misclassification penalty set twice that of malignant patients.

**Validation Testing:** The validation of the MPDL was done on an independent clinical data set (University of California-San Francisco-UCSF) was excellent[18,19]. **Figure 6** illustrates representative cases comparing the segmented tissue regions of the UCSF validation data set using the MPDL tissue signatures. The lesion volumes defined by MPDL and USCF resulted in a small percent difference of 4.4%±3.9%. Bland-Altman plots are shown in **Figure 7**, demonstrating no bias and excellent agreement between the data set. Finally, the sensitivity and specificity for differentiation of malignant from benign lesions were 90% and 85% respectively with an AUC of 0.93 and shown in **Figure 8**.

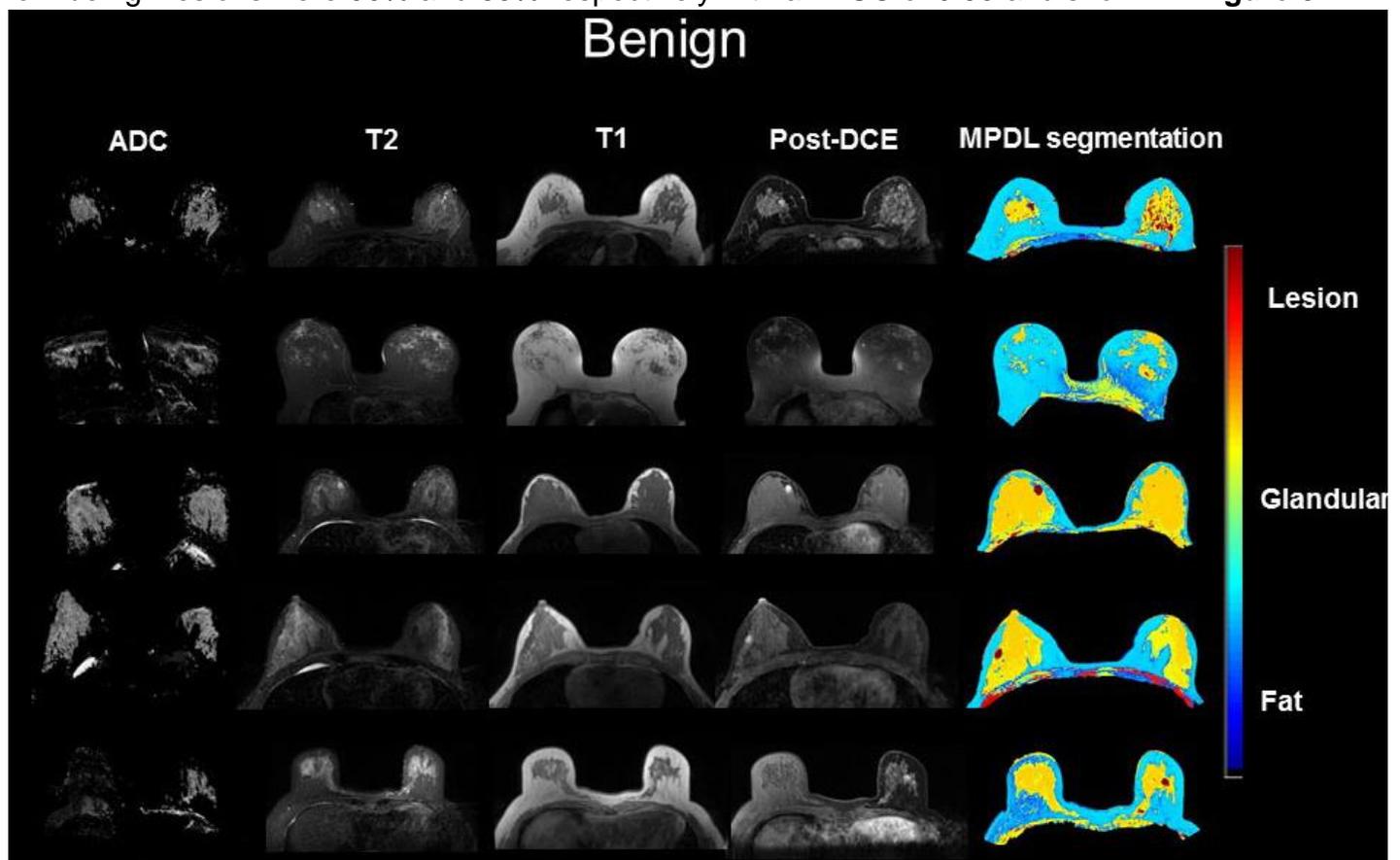

**Figure 4.** *Illustrates the use of MPDL network on axial breast mpMRI in five representative benign patients. The color coding for different tissue types are shown to the right of the images.*

## Discussion

Utilizing multiparametric deep learning network, we have developed, tested, and validated a cognitive computing platform that organizes, integrates, and interprets imaging information using a MPDL tissue signature model. The application of the MPDL tissue signature model resulted in excellent segmentation and classification of different tissue classes. This study employed an integrated multiparametric breast MRI deep learning model in the clinical setting and demonstrates that MPDL tissue signatures defines benign and malignant tissue and performs accurate classification. Moreover, this report demonstrates that deep learning-assisted unsupervised segmentation using mpMRI signatures can detect heterogeneous zones within breast lesions. These heterogeneous regions can be used for further classification of breast tissue by quantitative ADC maps and/or PK-DCE parameters. Finally, the MPDL model with machine learning classification distinguished between benign and malignant tissue with high sensitivity, specificity, and accuracy.

The results from application of MPDL tissue signature model to an independent breast MRI dataset demonstrated the robustness of the MPDL model. Importantly, the MPDL model was able to accurately segment breast tissue irrespective of the magnetic field strength (3T for our data and 1.5T for the validation set). Furthermore, the MPDL model was invariant to the imaging orientation as our dataset was in the axial plane while the validation set was obtained in sagittal plane. This invariance is due to the underlying depiction of the tissue using tissue signature vectors, which captures the tissue underlying characteristics and allows for the "adjustment" to different MRI input. Moreover, the MRI parameters, as well as, the time resolution of the DCE image phases used to train the MPDL model were different for our dataset and the validation dataset reasserting the robust nature of the MPDL model and eliminates the need to "retrain" the MPDL model.

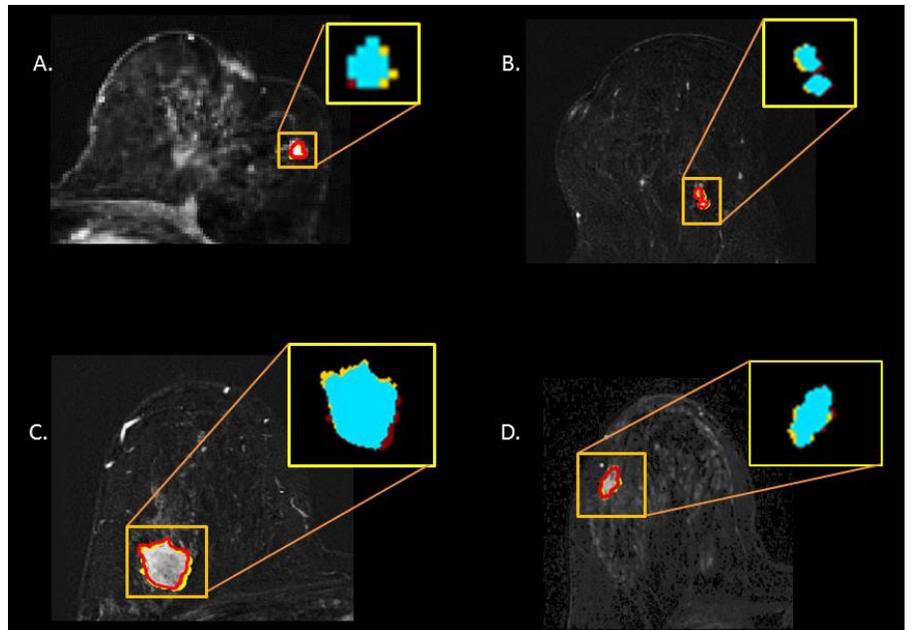

The mpMRI parameters used in this study were based on our and others previous results in patients [1,3,4,30]. These studies demonstrated that the combined MRI sequences consisting of DWI, ADC, and PK-DCE were highly correlated with the histological phenotype of the tissue. The sensitivity and specificity of classification between malignant and benign tumors by MPDL were similar to radiologists [31,32]. This is very encouraging when the future reading rooms will have advanced computing power to assist in reading of cases. Currently, it is very unlikely that machine learning and deep learning will replace radiologists as has been suggested by some, yet there may be a role for improved efficiency in the workflow and accuracy of interpretation. Using advanced computational methods with allow for this coming change to be better managed within radiology. Our results demonstrate that the MPDL method can be used on an independent data set acquired from different institutions. Indeed, the I-Spy trial is one of the largest MRI trials and incorporates many different MRI field strengths. However, the ability for the MPDL to learn different tissue signatures allows it to adapt to different data sets with highly accurate results. This was shown with the high dice similarity of the validation data using different input MRI data.

Figure 5. *Demonstration of dice similarity overlap between the Eigenimage and MPDL segmentation masks of two benign (A and B) and two malignant (C and D) patients overlaid on the subtracted dynamic contrast enhanced image. The Eigenfilter segmentation boundary is shown in yellow and the MPDL segmentation boundary is displayed in red. On the overlap masks, the blue region corresponds to the overlap between the two methods, red represents the area segmented by the MPDL alone while yellow represents the area segmented by Eigenfilter alone.*

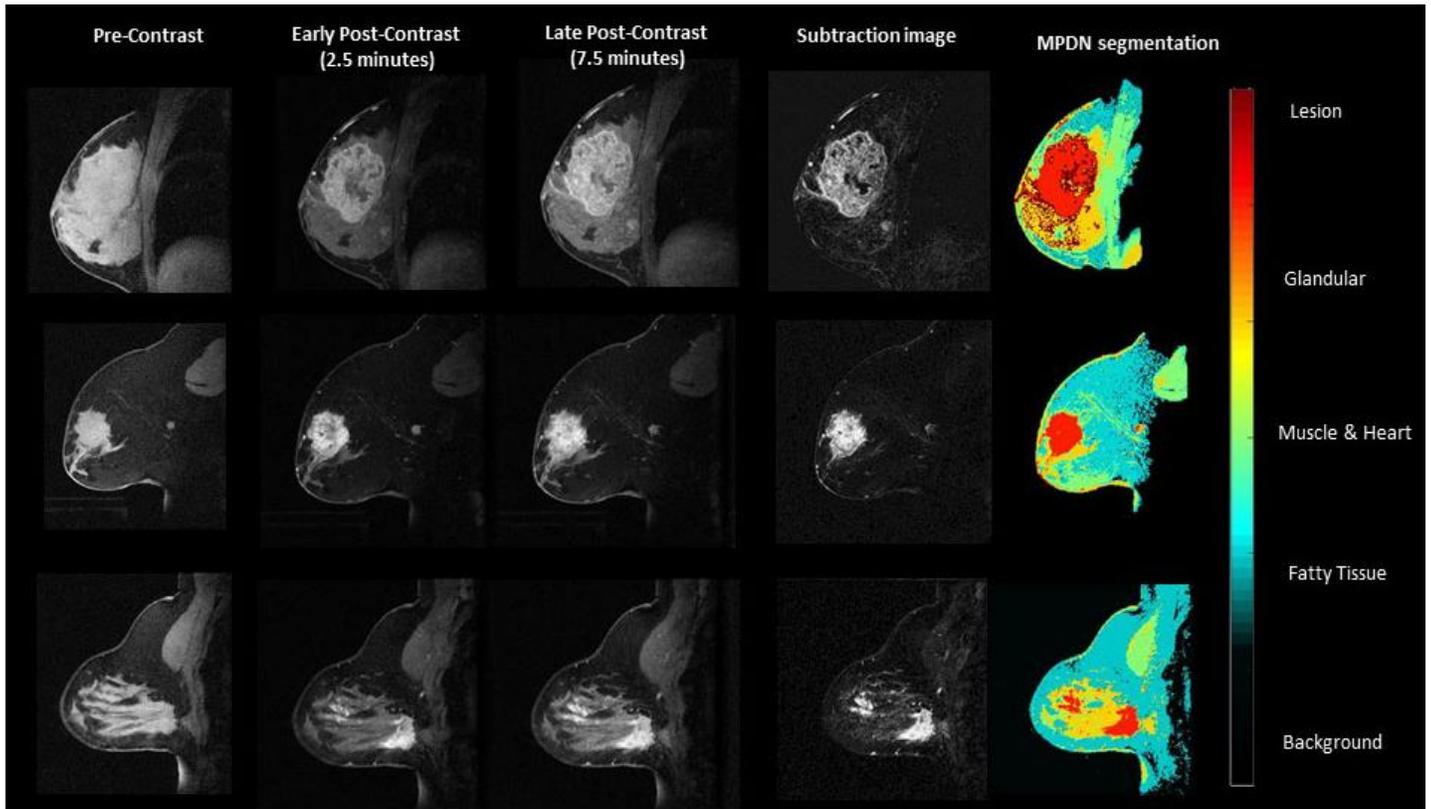

**Figure 6**. *Demonstration of three representative sagittal breast cases from the validation cohort and the resulting MPDL segmentations. In all cases the segmented regions of breast lesions were highly correlated between each other. The color coding for different tissue types are shown to the right of the images.*

There are, however, some technical limitations to the use the MPDL network in practice. First, increased computational power on the graphical processor units ((GPU) > 2500 cores, 12GB)) used here may not be widely available. However, the use of advanced GPU computing is rapidly finding applications in many different radiological datasets[10,12,15,33-36]. More specific to the present study, any assessment of the clinical value of MPDL network will require additional studies in a larger patient population. Moreover, a prospective trial with subsequent follow-up and pathological correlation using MPDL will provide us with new data to explore the exact application and methods to apply to larger studies.

In conclusion, we have demonstrated that integrated MPDL method accurately segmented and classified different breast tissue from multiparametric breast MRI. The MPDL images allow for improved visualization of different tissue characteristics based on multiple radiological parameters.

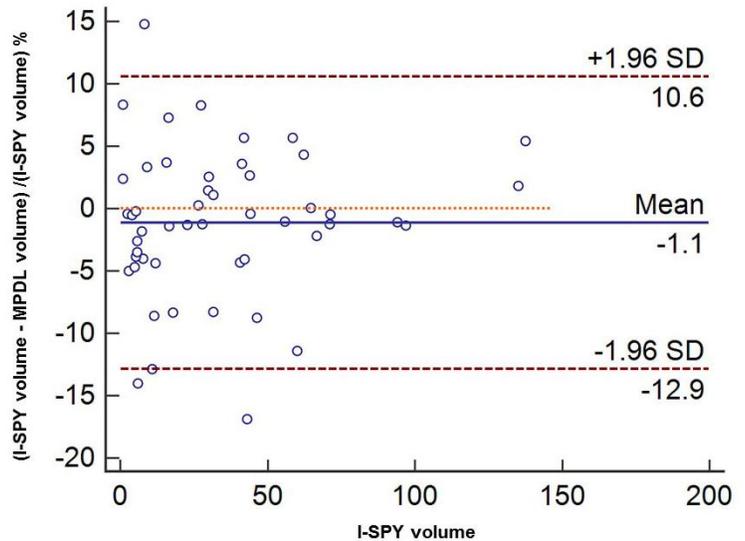

**Figure 7.** *Bland-Altman plots demonstrating the limits of agreement of the percent differences from the validation data set and MPDL segmentations. The mean is shown by the center line and the confidential intervals (±2SD) are shown at 10.6% and -12.9%. the resulting. The plot shows excellent agreement between the two measurements.*

**Code availability**
Our software will be freely available to academic users after issue of pending patents and a materials research agreement is obtained from the university. Due to University regulations, any patent pending software is not available until a patent is issued.

**Data availability**

All relevant clinical data are available upon request with adherence to HIPPA laws and the institutions IRB policies.

**Acknowledgments:** National Institutes of Health (NIH) grant numbers: 5P30CA006973 (Imaging Response Assessment Team - IRAT), U01CA140204, 1R01CA190299, and The Tesla K40s used for this research was donated by the NVIDIA Corporation.

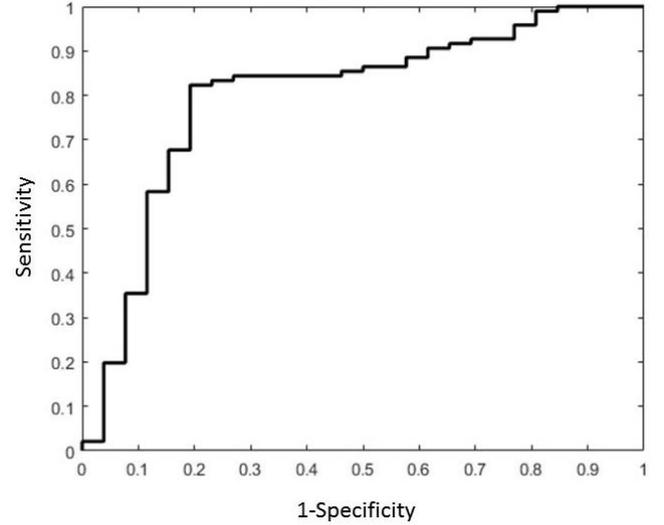

**Figure 8.** *The receiver operating characteristic curve from the MPDL classification shows an AUC=0.93 with a sensitivity of 90% and specificity of 85%.*

**Author Contributions:** MAJ and VP developed the concept, algorithm, and performed the testing, statistical methods, and manuscript writing and review. DB, IK, KM, RE, SH performed the data analysis and manuscript writing and review.
**Conflict of Interest:** The authors have no conflict of interests.

## Appendix

The MPDL framework is based autoencoders as shown in figure 2. An autoencoder has two parts: an encoder and a decoder. The encoder maps the vector X to the vector h$^{(1)}$ representing the hidden layer as follows:

$$\boldsymbol{h^{(1)}} = \boldsymbol{f}\,(\boldsymbol{WX} + \boldsymbol{b})$$

where f is the transfer function for the encoder, $W \in R^{d \times D}$ is the matrix of weights and b is the bias vector.
Where the decoder maps h$^{(1)}$ back to X using the following equation:

$$\widehat{\boldsymbol{X}} = \boldsymbol{g}(\boldsymbol{W'}\boldsymbol{h^{(1)}} + \boldsymbol{b'})$$

The values of W' and b' are equal to the transpose of W and b in case the weights are tied between encoder and decoder. The network comprises of layers and nodes within each layer. The nodes in the hidden layer were further specialized to activate in response to only a subset of the total number of MPDL tissue signatures using sparsity regularization. For example, after training the sparse autoencoder on a mpMRI dataset, node 1 may have "specialized" in activating only in response to a fatty tissue signature while node 2 may have "specialized" in activating only in response to a glandular tissue signature. Mathematically, the average activation, $\widehat{\rho}_j$ of a neuron, j is given by the following equation

$$\widehat{\rho}_j = \frac{1}{N}\sum_{i=1}^{N} h_j^{(1)}(x_i)$$

where N is the total number of training samples. If $\rho$ denotes the desired average activation or the sparsity proportion of the neuron, j across all the training samples, our goal is to impose the constraint $\widehat{\rho}_j = \rho$. Consequently, the sparsity regularization term added to the cost function is given as

$$R_S = \beta \sum_{j=1}^{d} \left( \rho \log \hat{\rho}_j + (1-\rho) \log \frac{1-\rho}{1-\hat{\rho}_j} \right)$$

where $\beta$ is the sparsity regularization penalty. Because we have four tissue classes, the sparsity proportion, $\rho$ was set at 0.25 and the penalty, $\beta$ was set at 4 to train each sparse autoencoder.

The output of the final sparse autoencoder was used as input to train a softmax classifier to classify the tissue signatures as background, fat, glandular or lesion. The cost function for training the softmax layer of the mp MRI segmentation deep network was based on cross entropy, given by

$$J = \frac{1}{N} \sum_{i=1}^{N} \sum_{j=1}^{4} t_{ij} \ln y_{ij} + (1 - t_{ij}) \ln(1 - y_{ij})$$

where $t_{ij}$ is the target class and $y_{ij}$ is the output of the deep network at the softmax classification layer. The weights of the pre-trained network were further fine-tuned using scaled conjugate gradient backpropagation method to improve the classification accuracy of the pre-trained network. The pre-trained weights are especially useful when the application is limited by the number of available training examples. After training, the first layer of the SSAE learns the most representative tissue signatures from the input dataset while the stack of subsequent autoencoders forms a composite representation of the tissue signatures, such that each node of the final layer specializes in recognizing tissue signatures from a single tissue type.

# Figures

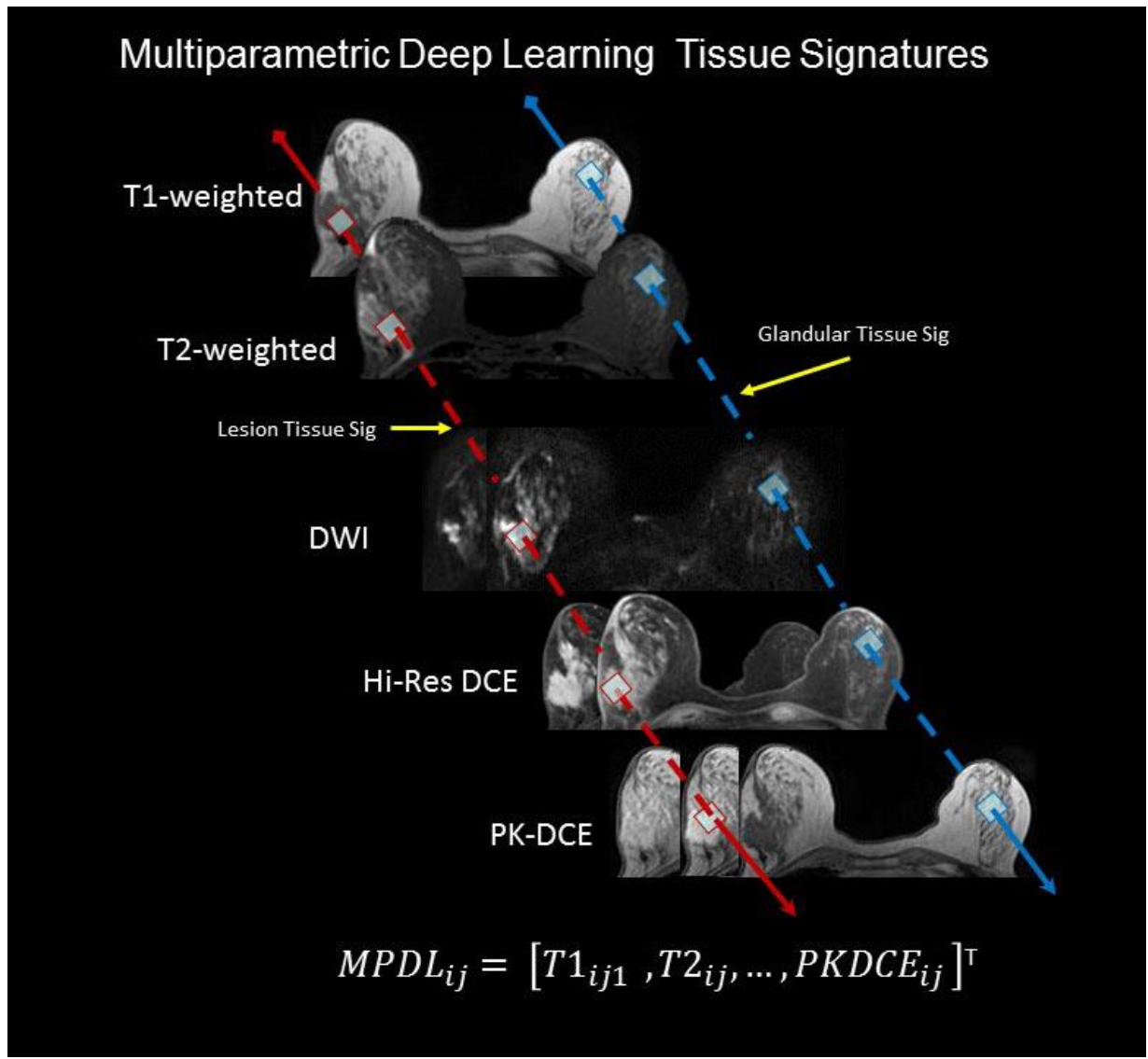

Figure 1. Demonstration of the multiparametric deep learning tissue signatures on axial mpMRI of the breast. Representative tissue signatures from the normal and abnormal tissue are obtained on each of the input MRI to create the vector signature MPDL$_{ij}$.

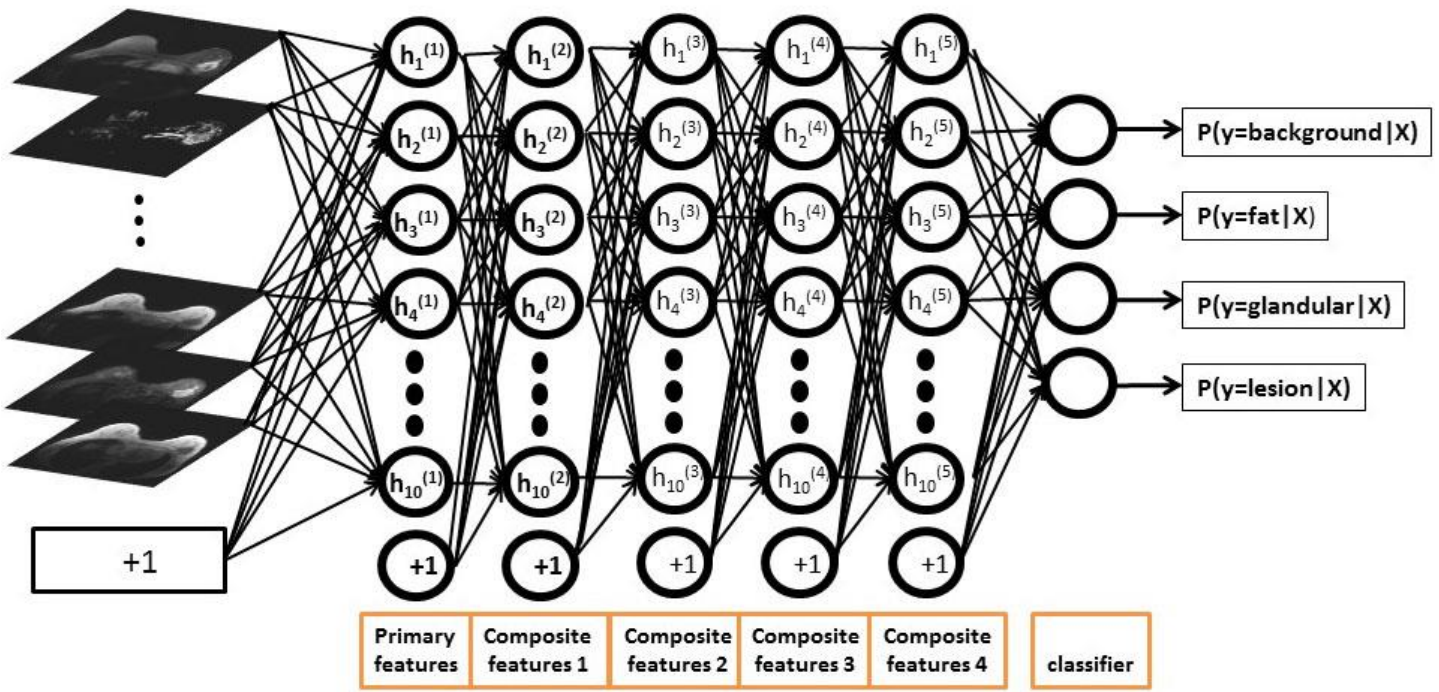

Figure 2. Framework for mpMRI segmentation deep network trained to segment multiparametric breast MRI into regions of different tissue types and background. The stacked sparse encoder deep network is constructed of ten hidden layers with five nodes each and a softmax classification layer that outputs the probability of different tissue types for the input tissue signature.

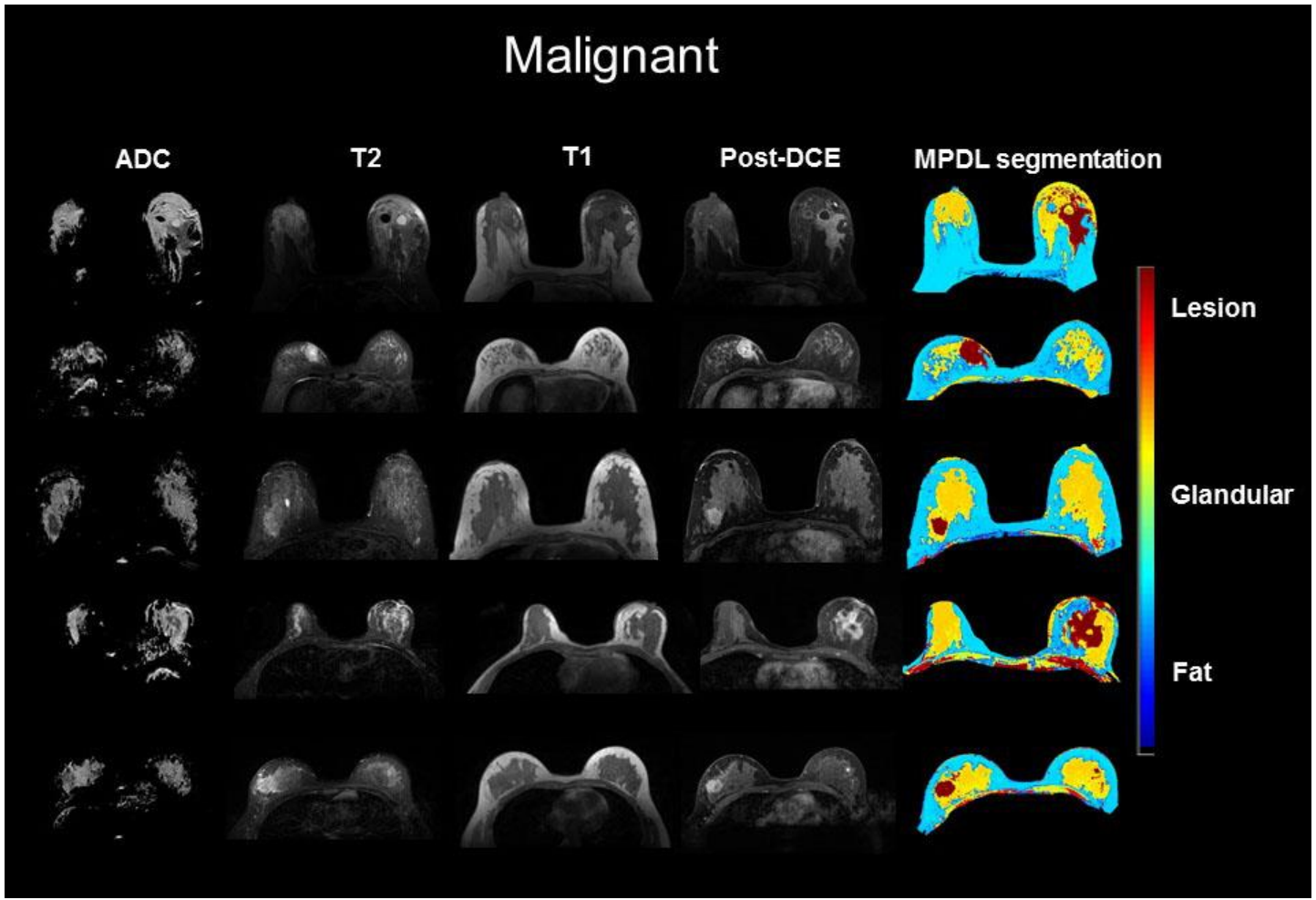

Figure 3. Illustrates the use of MPDL network on axial breast mpMRI in five representative malignant patients. The color coding for different tissue types are shown to the right of the images.

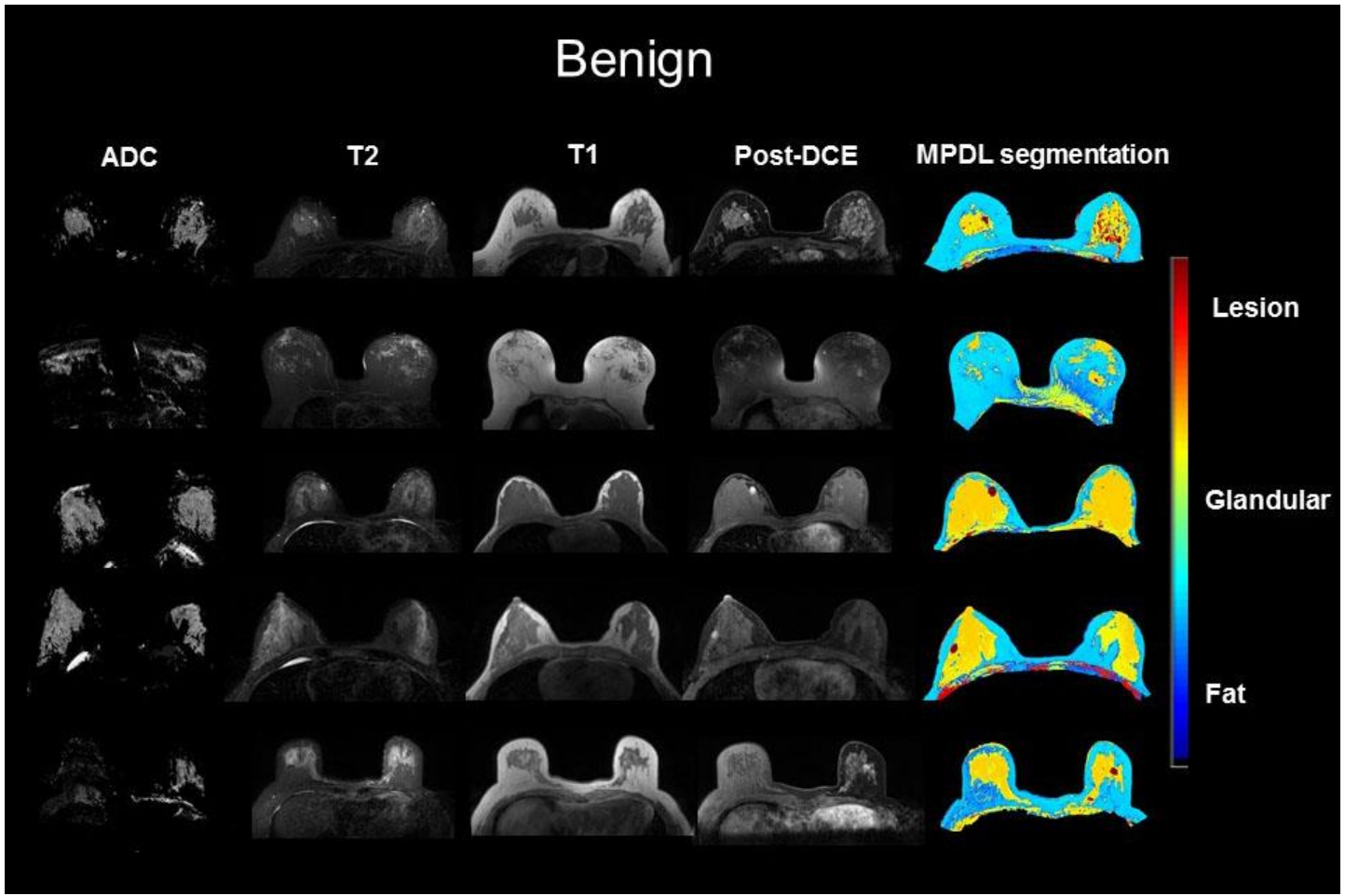

Figure 4. Illustrates the use of MPDL network on axial breast mpMRI in five representative benign patients. The color coding for different tissue types are shown to the right of the images.

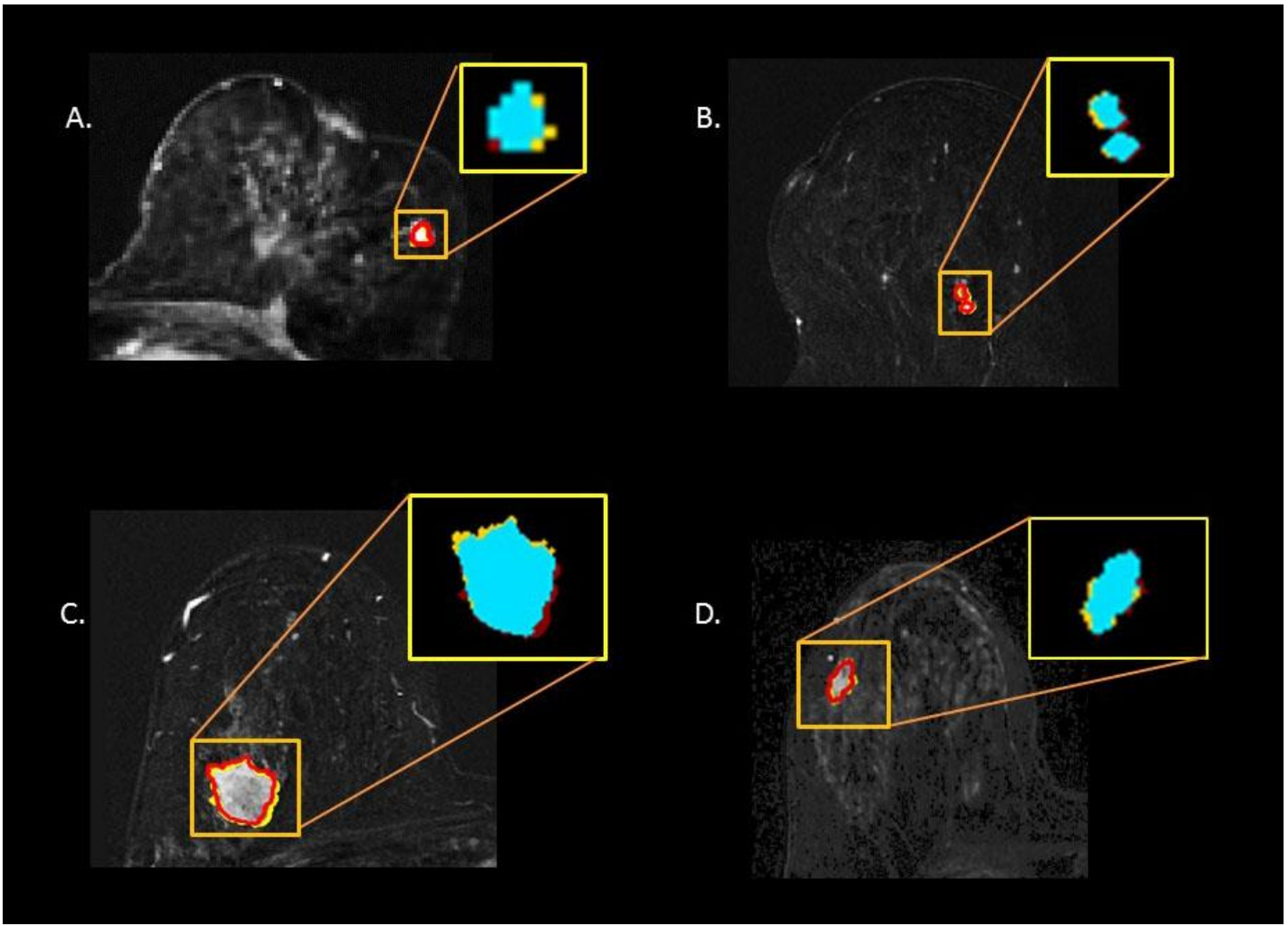

Figure 5. Demonstration of dice similarity overlap between Eigenimage and MPDL segmentation masks of two benign (A and B) and two malignant (C and D) patients overlaid on the subtracted dynamic contrast enhanced image. The Eigenfilter segmentation boundary is shown in yellow and the MPDL segmentation boundary is displayed in red. On the overlap masks, the blue region corresponds to the overlap between the two methods, red represents the area segmented by the MPDL alone while yellow represents the area segmented by Eigenfilter alone.

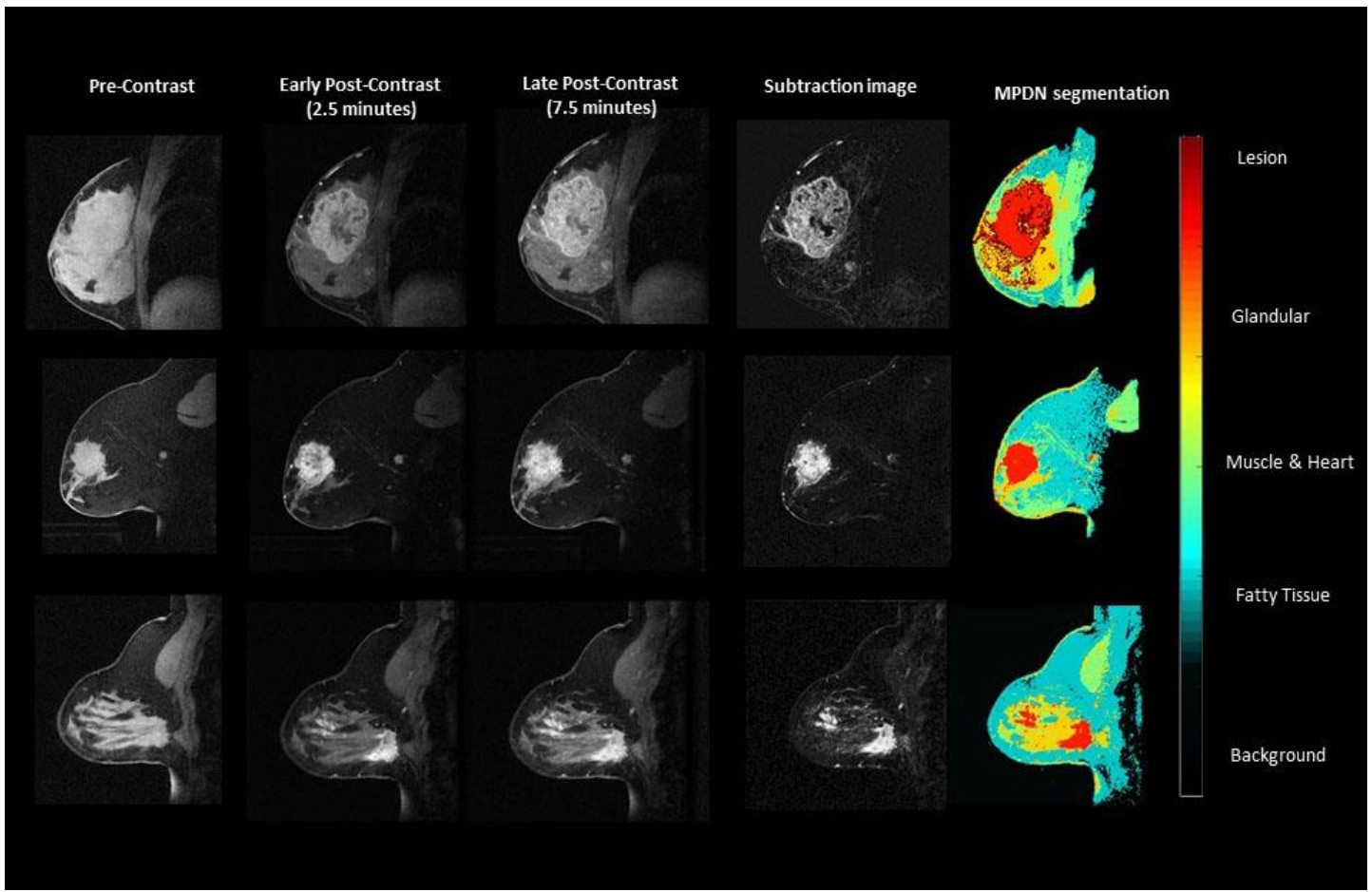

Figure 6. Demonstration of three representative sagittal breast cases from the validation cohort and the resulting MPDL segmentations.  In all cases the segmented regions of breast lesions were highly correlated between each other. The color coding for different tissue types are shown to the right of the images.

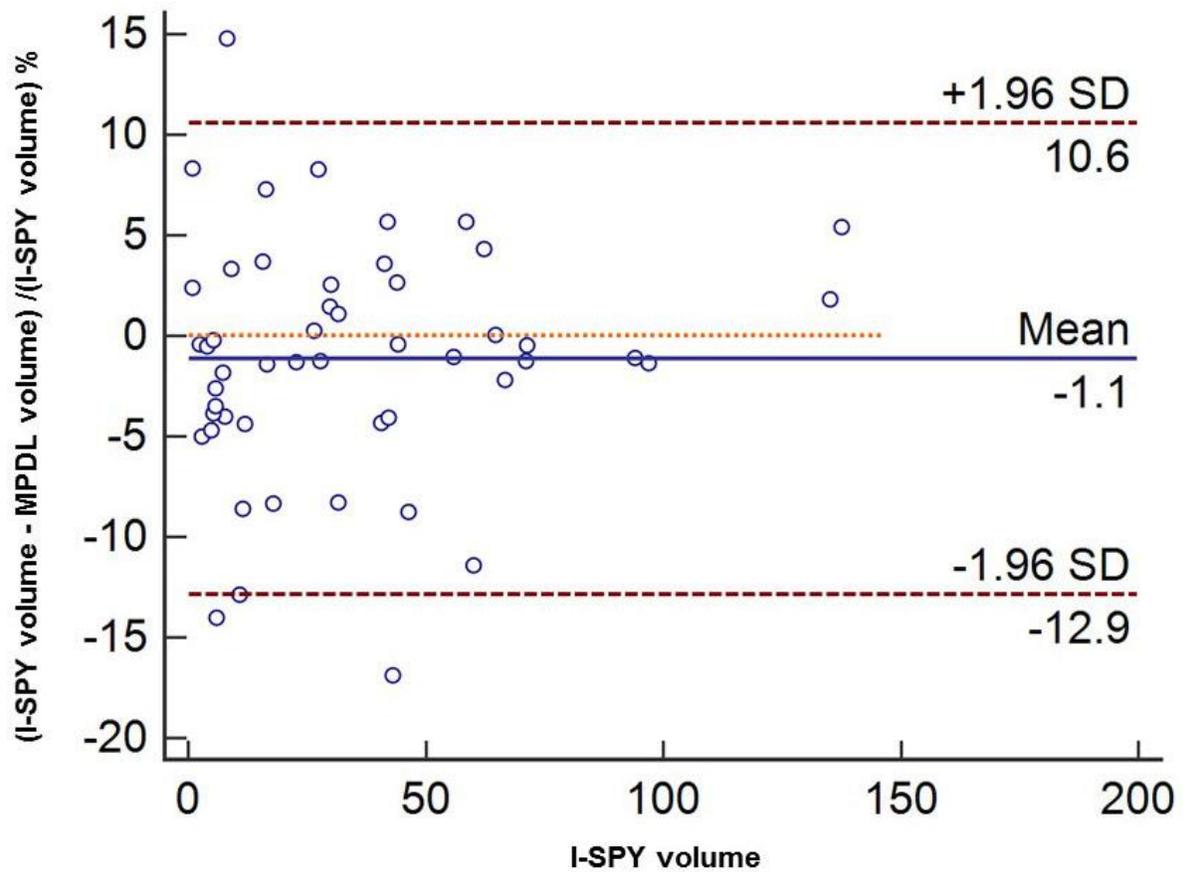

Figure 7. Bland-Altman plots demonstrating the limits of agreement of the percent differences on the representative sagittal breast cases from the validation data set and MPDL segmentations. The mean is shown by the center line and the confidential intervals (±2SD) are shown at 10.6% and -12.9%. the resulting. The plot shows excellent agreement between the two measurements.

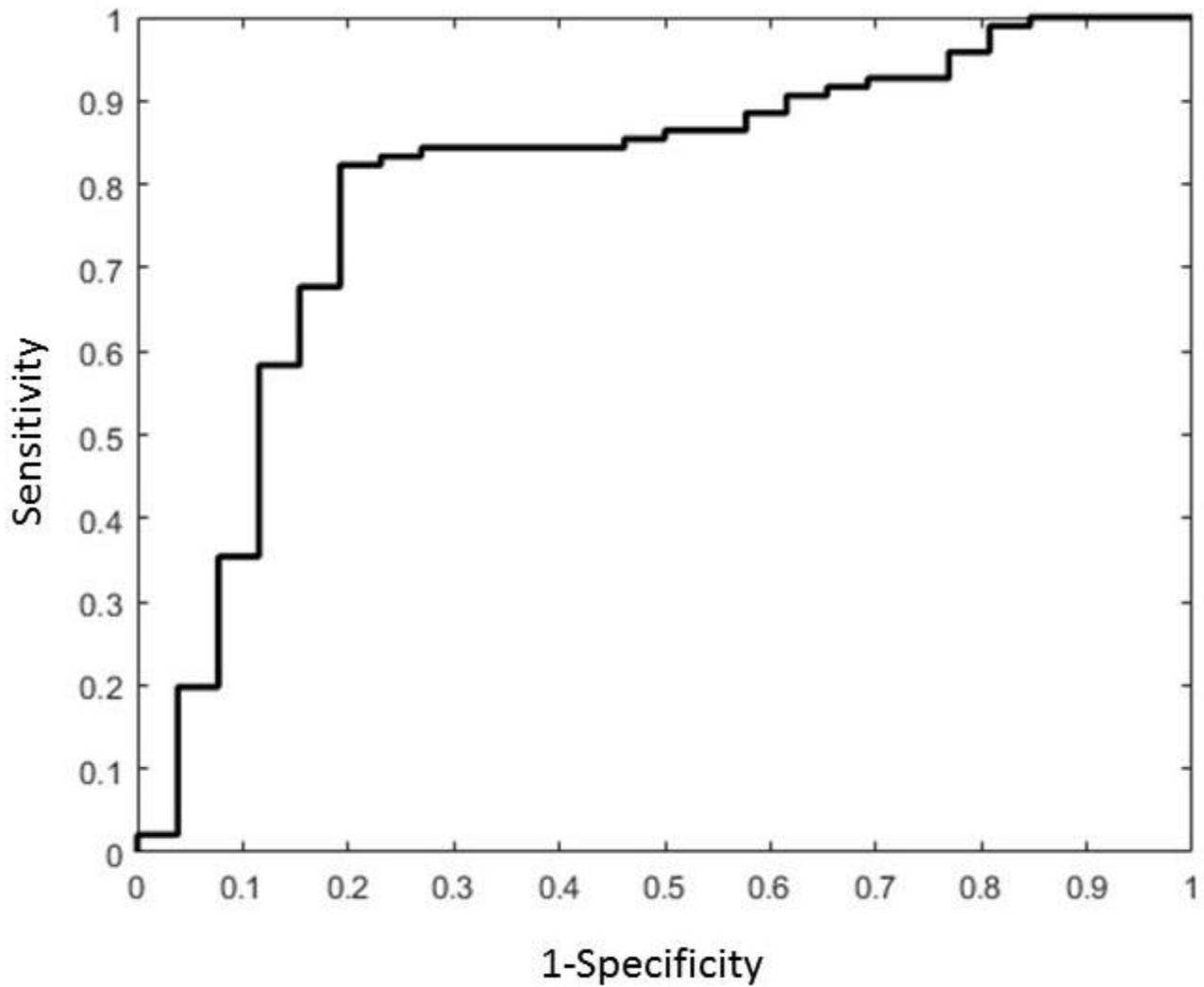

Figure 8. The receiver operating characteristic curve from the MPDL classification shows an AUC=0.93 with a sensitivity of 90% and specificity of 85%.